\documentclass[a4paper]{jpconf}
\usepackage{graphicx}
\begin{document}
\title{Theoretical and experimental activities on opacities for a good interpretation of seismic stellar probes}
\author{ S Turck-Chi\`eze$^1$, G Loisel$^{1, 2}$, D Gilles$^1$, F Thais$^2$, S Bastiani$^3$, C\,Blancard$^4$, M\,Busquet$^5$, T\,Caillaud$^4$, P\,Cosse$^4$, T\,Blenski$^2$, F\,Delahaye$^6$, J\,E\,Ducret$^9$, G\,Faussurier$^4$, F\,Gilleron$^4$, J\,Guzik$^7$, J\,W\,Harris$^8$, D\,P\,Kilcrease$^7$, N\,H\,Magee$^7$, L\,Piau$^1$, J\,C\,Pain$^4$, M\,Poirier$^2$, Q\,Porcherot$^4$, C\,Reverdin$^4$, V\,Silvert$^4$, B\,Villette$^4$ and C\,Zeippen$^6$ }
\address{$^1$IRFU/ CEA/ CE Saclay, F-91190 Gif sur Yvette, $^2$ IRAMIS/ CEA/ CE Saclay, F-91190 Gif sur Yvette, $^3$LULI, Ecole Polytechnique, CNRS, CEA, UPMC, F-91128 Palaiseau Cedex, $^4$CEA DAM/DIF, F-91297 Arpajon, $^5$ARTEP, Ellicott City, MD 21042, USA, $^6$LERMA and LUTH, Observatoire de Paris, ENS, UPMC, UCP, CNRS, 5 Place Jules Janssen, F-92195 Meudon, $^7$Theoretical Division LANL, Los Alamos NM 87545,USA, $^8$AWE, Reading, Berkshire, RG7 4PR, UK,$^9$ CELIA, Bordeaux, France }
\ead{sylvaine.turck-chieze@cea.fr}
\begin{abstract}
Opacity calculations are basic ingredients of stellar modelling. They play a crucial role in the  interpretation of acoustic modes detected by SoHO, COROT and KEPLER. In this review we present our activities on both theoretical and experimental sides. We show new calculations of opacity  spectra and comparisons between eight groups who produce  opacity spectra calculations in the domain where experiments are scheduled. Real differences are noticed with real astrophysical consequences when one extends helioseismology to cluster studies of different compositions. Two cases are considered presently: (1) the solar radiative zone and  (2) the beta Cephei envelops. We describe how our experiments are performed and new preliminary results on nickel obtained in the campaign 2010 at LULI 2000 at Polytechnique.
\end{abstract}

\section{Introduction}
Opacity coefficients are basic elements of stellar equations like equation of state and reaction rates in the  stellar zones where the radiative gradient is smaller than the adiabatic one, that means the regions where the transport of energy is dominated by the photon interaction with matter. These coefficients, $\kappa(T, \rho, X_i)$ expressed in cm$^2$/g, represent the interacting cross sec- tions of  photon with matter. They are calculated at each mesh of a stellar model in radius and time and depend strongly on the temperature and composition but less on the density.
In stellar interiors, these interactions are considered in Local Thermodynamic Equilibrium and quasi instantaneous. This last point could appear less justified than the first one due to the stochastic displacement of photons but stays a reasonable approximation to follow the great stages of evolution which  describe the radiative transport of energy.   
It was established several decades ago, that these cross sections correspond to  the Rosseland mean values of the corresponding photon spectrum within the diffusion approximation (Cox \& Giuli, 1968; Clayton 1983).

The corresponding complex spectra must contain all the different processes that the different constituants of the plasma experience and they strongly depend on the degree of ionization of each species.  It is known that these spectra result from a well knowledge of atomic and plasma physics. Their production needs a dedicated huge work which has been mainly done by Los Alamos (Huebner et al. 1977) and Livermore (Iglesias, Rogers \& Wilson, 1987) groups producing tables for the astrophysical uses. Then the completude of elements and a lot of corrections have been introduced in the OPAL tables of Livermore (Iglesias \& Rogers 1996) following also some comparisons with first experiments. During the last decade,  another team has produced OP tables and spectra for astrophysical application (Seaton 2005 and references therein). It is important to recall that the opacity spectra are not only useful to describe, through the mean Rosseland values, the transport of energy, they are also important for the estimate of the radiative acceleration along the stellar lifetime and for the prediction and interpretation of the stellar acoustic modes (Turck-Chi\`eze et al. 2009).

Today with the success of global helioseismology (Turck-Chi\`eze et al. 1993, Thompson et al. 1996, Turck-Chi\`eze  et al. 2001, Basu et al. 2009) and the development of asteroseismology (Aerts, Christensen-Dalsgaard \& Kurtz 2010), it is the appropriate time to focus more deeply on this fundamental ingredient of stellar evolution to properly identify and interpret the present and future space missions.  In this review, we describe the new activities developed by our team to give credit to the used opacity calculations.  

In Turck-Chi\`eze et al. (2009), we point out two cases which must be examined in great details:  (1) the radiative zone of Sun and solar like stars, where the observed sound speed (Turck-Chi\`eze et al. 2001, Basu et al. 2009, Turck-Chi\`eze \& Couvidat 2010) is not yet understood, (2) the envelop of $\beta$ Cephei where the excitation of the oscillations is due to the opacity peak of the iron group and for which there is some well identified difficulty of interpretation (Pamyatnykh 1999). To progress on these two fields we have formed a consortium between plasma and astro physicists in order to compare calculations and to perform experiments on high energy lasers.

\section{The opacity coefficients in radiative zones of Sun and solar-like stars}
Figure 1 illustrates the successive role of the heavy elements in the increase of the opacity cross section of the solar radiative zone along radius (Turck-Chi\`eze et al. 1993). In fact the central region is sufficiently hot for all the elements to be completely ionized except for the elements in Z equal or greater than iron. Then  one observes the role of lighter and lighter elements from the center to the base of the convective zone. A radiative gradient greater than the adiabatic one results from the strong increase of opacity due to the partial ionization of oxygen (the third element in abundance). When this phenomenon appears,  the convective transport takes over. Just below the photosphere (the zoom at the right part of the figure), helium then hydrogen dominates successively with bound-bound and bound-free contributions in addition to  free-free or diffusion ones.
\begin{figure}
\begin{center}
\includegraphics[width=16pc]{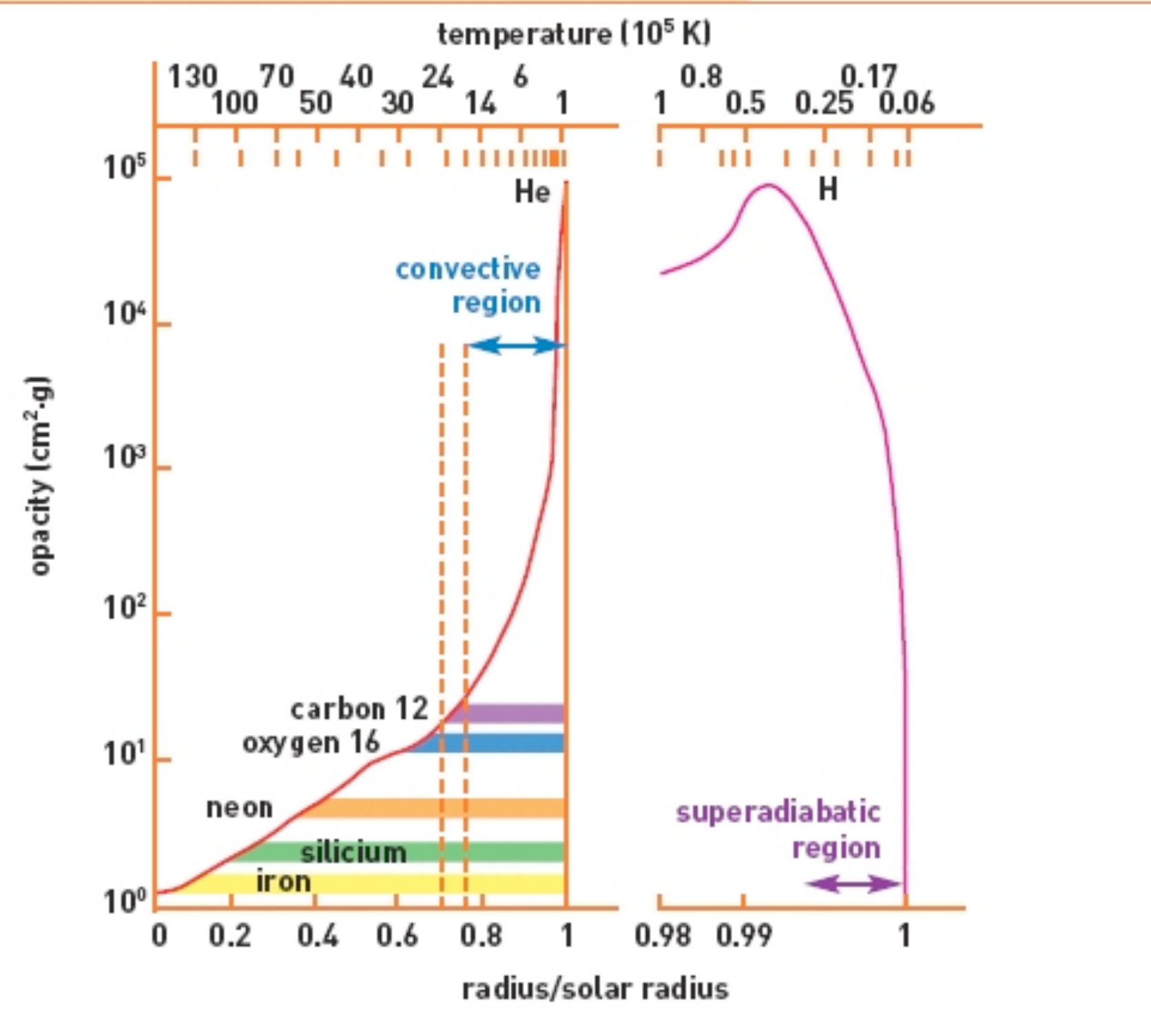}\hspace{2pc}
\end{center}
\caption{\label{trans} Main heavy elements contributors to the opacity of the radiative zone of the Sun along the solar radius. From Turck-Chi\`eze et al. 1993. }
\end{figure}
This property of the solar opacity coefficients is known for a long time and by chance the different calculations, OPAL and OP, agree reasonably well in Rosseland mean values (Seaton \& Badnell 2004) within 5\%. Nevertheless, these coefficients depend strongly on the detailed composition and more specifically on O, Ne, and Fe contributions. 

Figure 2 shows, on the contrary, that the agreement on elemental Rosseland values is not so good for most  of the heavy elements comparing OP and OPAS calculations. Just below the base of the convective zone and down to 0.5 R$_\odot$, differences, up to 60\% for some specific elements, are noticed between OP and OPAS. OPAS are new  calculations of opacity dedicated to stellar applications including 21 elements  and performed by a new team in CEA (Blancard, Coss\'e and Faussurier 2010). The noticed  differences are important for the treatment of the radiative acceleration in microscopic diffusion which uses the individual spectra (see Turck-Chi\`eze et al. 2009, for the different expressions used in stellar evolution for treating the different radiative effects).

\begin{figure}
\begin{center}
\includegraphics[width=25pc]{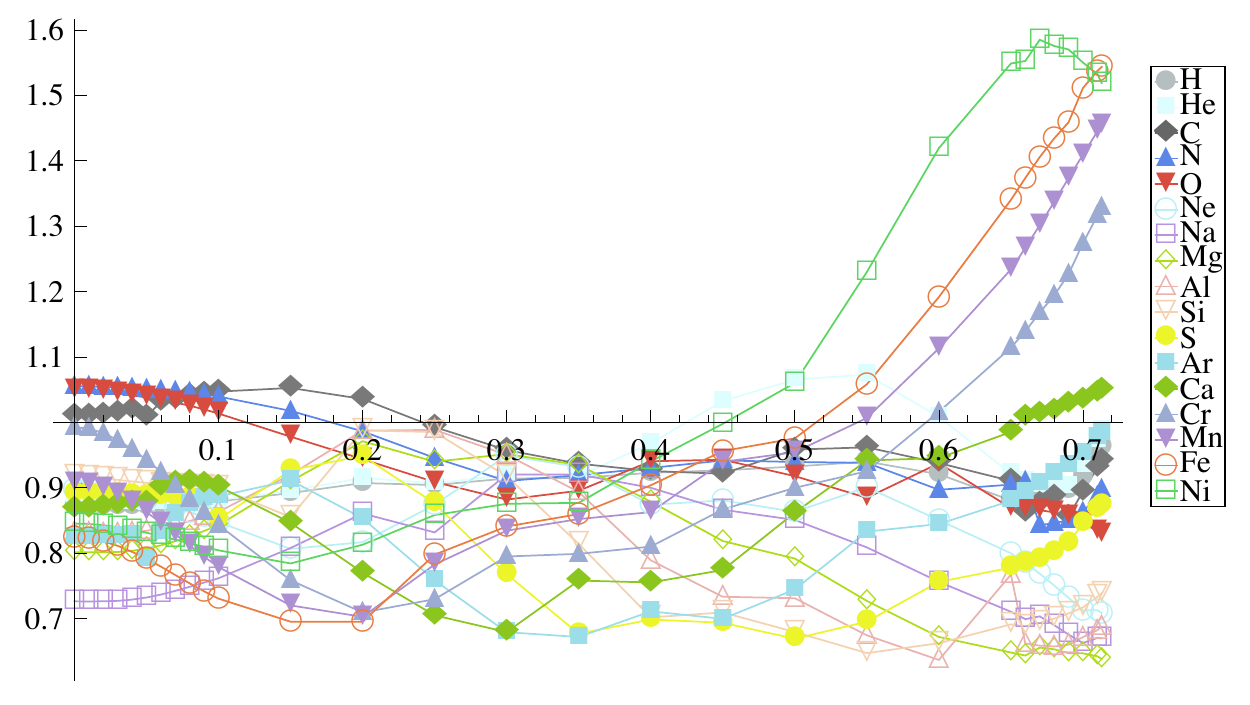}
\end{center}
\caption{\label{trans} Difference versus solar radius between OPAS and OP Rosseland mean values for each individual contributor to the solar opacity.  From Blancard, Coss\'e and Faussurier 2010.}
\end{figure}
The present situation illustrated by figures 1 and 2  is not totally surprising because the total Rosseland mean value integrates contribution  of a lot of elements and  depends strongly on the free-free process of helium, the bound-free of iron and the position of the main bound-bound lines of the partially ionized elements. On the opposite, the individual mean Rosseland value for each element strongly depends on the specific lines considered for each element, their width  and the Stark effect, all strongly dependent on the approach used.  It is why in addition to check the validity of the calculations used  it is important  to perform  also experiments which reveal some aspects of the spectra.
The solar radiative conditions are not yet been checked but a first experiment has been realized in the Z pulsed power facility of  Sandia lab. in Albuquerque at 1.8 10$^6$ K and an electron density about 100 times smaller than the base of the convective zone which is in reasonable agreement with OPAL and OPAS estimate of the iron and magnesium spectra (Bailey et al. 2007, 2009).

\section{Preliminary results on Nickel for envelops of  $\beta$ Cephei}
The  $\beta$ Cephei stars (8$\; < M < 12 M_\odot$) pulsate by $\kappa$ mechanism due to a strong peak of opacity of the iron group. Unfortunately the two available opacity tables OP and OPAL lead to strong differences in mean Rosseland values for the four contributors Cr, Mn, Fe and Ni (see Turck-Chi\`eze 2010). This fact leads to the difficulty to choose which  table is the best to use  for interpreting the corresponding pulsations observations (Daszynska-Daszkiewicz  \& Walczak 2010, Degroote et al. 2009). So in order to better understand these differences we decided to perform an experiment on these elements and impulse some comparison between codes. 

Unfortunately, it is not possible to perform an experiment at the  too low densities of these envelops so we have determined equivalent conditions of plasma where the degree of ionization is rather similar. We decided also to check two or three conditions of temperature because the opacity spectra change quickly with temperature. We have shown  how the iron spectrum itself can differ between the 8 calculations which are participating to the comparison (see Turck-Chi\`eze et al., 2010 for the spectra and a rapid code description).  So, one may hope a good discrimination from the experiment. Figure 3 shows different calculations for the nickel spectrum in two bands of wavelengths corresponding to the experiences we have performed, it appears clearly that OP calculations differ strongly from the other calculations for temperature around 15 eV (170 000 K). 
Nickel, contrary to iron, has never been measured in the wavelength range corresponding to the mean Rosseland value at such temperature. In the following,  we report on the first nickel spectrum measurements done this year by our consortium. 
 \begin{figure}
\begin{center}
\includegraphics[width=18pc]{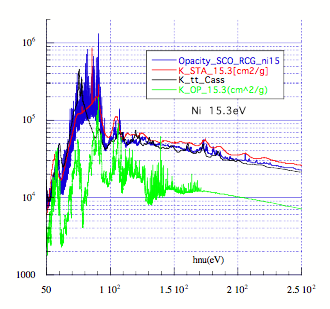}\hspace{2pc}
\includegraphics[width=16pc]{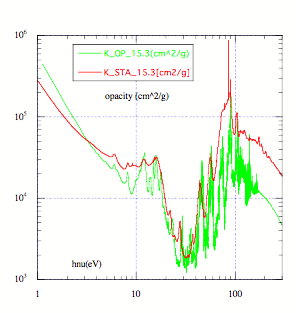}
\end{center}
\caption{\label{trans} Theoretical opacity nickel spectra corresponding to 15.3 eV and 5.7 10$^{-3}$ g/cm$^3$ and obtained from different approaches : OP (Seaton \& Badnell, 2004; Seaton, 2007), Cassandra (Crowley et al. 2001), STA (Bar-Shalom, 1989) and SCO-RCG (Porcherot et al., 2010)  in two wavelength ranges:  the left one corresponds to the range that must be explored to calculate the mean Rosseland value, the right one corresponds to a zoom near the maximum at 62 eV of the Rosseland mean value.}
\end{figure}

The spectral opacity measurements have been performed at LULI 2000 with  two complementary lasers (see www.luli.polytechnique.fr). 
 A nanosecond laser delivers an energy between 30-500 J in a 500 ps duration pulse. This laser is used to irradiate a gold cavity (hohlraum) on which a foil of the considered element  (here nickel) is deposited and heated. After a delay chosen to get the required density and temperature (this delay is determined by a simulation of the geometry and heating of the experiment), a picosecond laser interacts with a backlighter foil to produce x rays in a short pulse (10-30ps)  to probe the formed plasma. 
 
 The transmission spectrum of the photons: $T (\nu)= \exp( - \kappa(\nu) \rho r)$ where r is the thickness of the foil,  is measured by a streak camera placed behind a specifically designed XUV-ray spectrometer. A detailed description of the experimental set up and of the experiment analysis will be published by Loisel et al. (2011).  The quality of these measurements  requires a sufficient spectral accuracy (about 1 eV) and a rather small emission of the cavity during the measurement of the spectrum. A detailed analysis of all the previous experiments and their required conditions can be found in Chenais-Popovics (2002), Bailey et al. (2009) and Loisel et al. (2009). 
 
 For the specific measurements that we have done on Cr, Ni, Fe, Cu and Ge, we need to form a plasma in LTE at the required conditions. This supposes first a good simulation of the whole experiment to probe the foil at the best moment to get the appropriate temperature without pollution of the gold of the cavity. Moreover, the rapid expansion of the foil during the heating is limited by placing the foil between two thin samples of  a low Z material here carbon. We measure  the foil temperature on one side thanks to  $\mu$DMX, a 12 channels spectrometer which measures x-rays energy (Bourgade et al., 2001) and we have limited the gradient of temperature below 10\% inside the foil in separating the incident ns beam in two parts and placed the foil between two cavities. The streak camera gives a time dependence of the phenomenon (resolution of about 50 ps) allowing to discriminate between backlight signal and self emission of the cavity.

\begin{figure}
\begin{center}
\includegraphics[width=16pc]{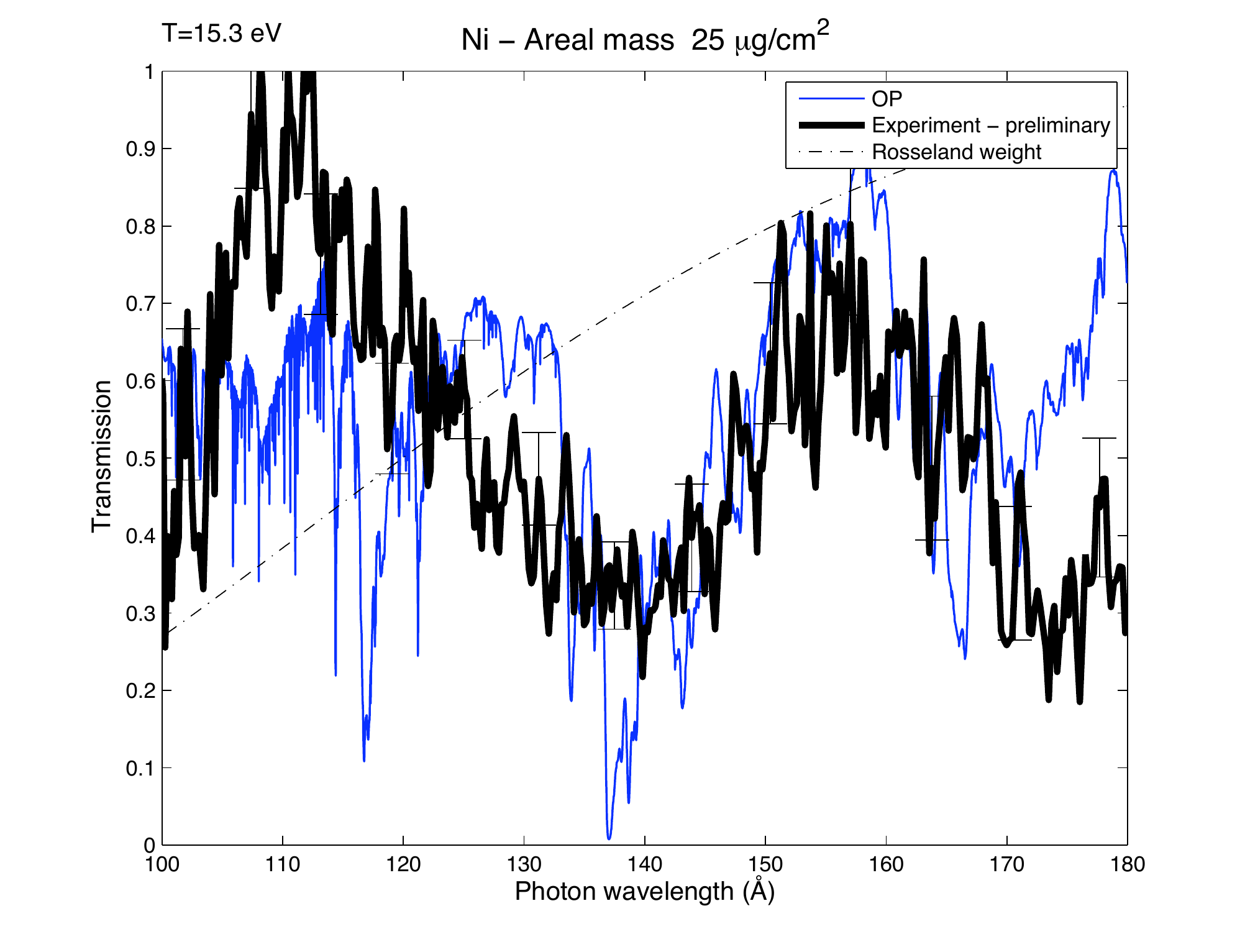}\hspace{2pc}
\includegraphics[width=16pc]{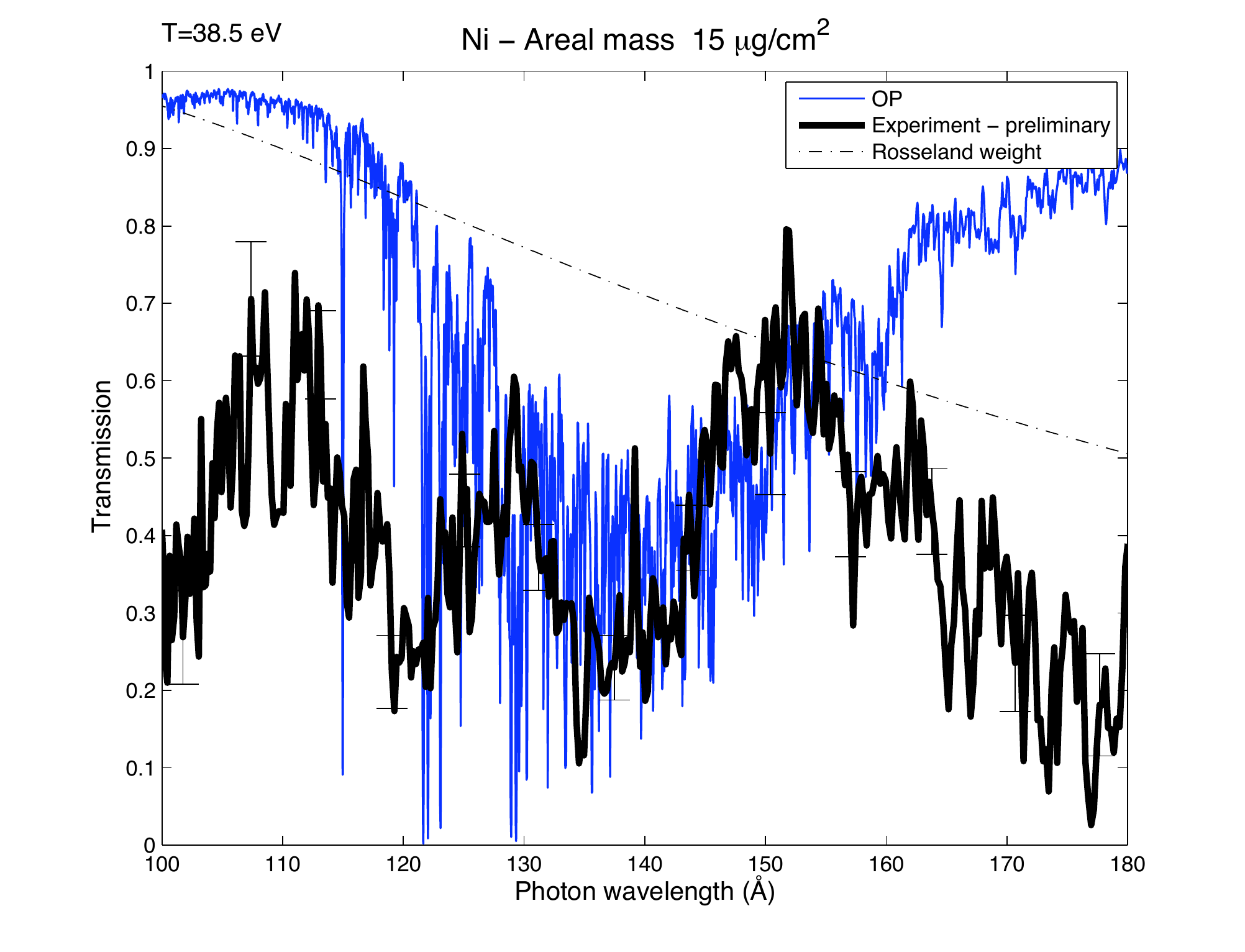}
\end{center}
\caption{\label{trans}Preliminary result of the transmission spectra obtained for nickel (in heavy black) measured at LULI 2000 compared to the OP calculation (blue) already shown in Figure 3 for two conditions of temperature (15.3 eV and 38 eV) and respectively $\rho$ = 2.7 mg/cm$^3$ Ne= 3.16 10$\rm ^{20}cm^{-3}$. Superimposed  is the weight of the Rosseland mean value normalized to 1  in dotted line. The statistical error bars are shown on the experimental spectra and the final comparison will integrate the temperature gradient in the foil and a proper estimate of the mean temperature.}
\end{figure} 
Figure 4 shows the transmission spectra obtained for nickel and compared to OP calculation for two conditions of temperature around the value which corresponds to the same degree of ionisation than in $\beta$ Cephei. We have also compared these spectra to the other calculations. Figure 4 does not present a perfect agreement between experiment and calculation, on  the whole range observed but the same comparison disagrees more strongly with the other calculations. We have already introduced in the observed spectra a statistical error bar which increases on the limit of the range due to the relative increase of the effect of the background. Systematic error may be still present and a better knowledge of the detector systematic effect is under study. It would be interesting to see if the other spectra on iron, chromium, copper and germanium show the same preference for OP calculation and to confirm such result in a more extensive study.

This review has shown through two examples that the interpretation of the seismic observations need to use appropriate opacity coefficients to extract without ambiguity  the manifestations of stellar dynamics. From  the existing experimental studies one cannot exclude that the different regions of stars require more effort on the theoretical calculations and that the present ones differ essentially by their approach which is better adapted for some terms and not some others due to the methods used (statistical and detailed configuration approaches or interaction between configurations). Moreover depending on their use, different efforts have been applied. Presently, we must also be cautious because the experiments are extremely complex, different approaches are applied for the covered wavelengths (X  or XUV techniques) for the machines used (Z machines or high energy lasers). Even the experimental approach exists since more than 10 years,  they need certainly to be repeated and extended in wavelength ranges and elements considered face to the development of asteroseismology.

\section*{References}
\begin{thereferences}
\item Aerts C, Christensen-Dalsgaard J \& Kurtz D W 2010 {\it Asteroseismology} A\&A edition, Springer
\item Asplund M, Grevesse N and Scott P 2009 {\it ARA\&A} {\bf 47} 481
\item Bailey J et al. 2007 {\it Phys. Rev. Lett.} {\bf 99} 5002
\item Bailey J et al. 2009 {\it Phys. of Plasmas} {\bf 16} 058101
\item Bar-Shalom, A., Oreg, J., Goldstein, W. H., Schwarts, D., Zigler, Z. 1989 {\it Phys. Rev. A}{\bf  40}  3183 
\item Basu S et al. 2009 {\it ApJ} {\bf 699} 1403
\item Blancard C, Coss\'e P \& Faussurier G 2010 {\it A\&A Suppl. Series} in preparation
\item Bourgade J L et al. 2001 {\it Rev. Sci. Inst} {\bf 72} 1173
\item Chenais-Popovics C 2002 {\it Laser and Particle Beams} {\bf  20} 291
\item Clayton D, 1983 {\it Principle of stellar evolution and nucleosynthesis} Mc Graw Hill
\item Cox J and Giuli R 1968 {\it Principles of stellar structure Applications to stars} Gordon and Breach
\item Crowley, B.J.B  et al.  2001 {\it JQRST} {\bf 71} 257
\item Daszynska-Daszkiewicz J \& Walczak P 2010 {\it MNRAS} {\bf 403} 496   
\item Degroote P et al. 2009 {\it A\&A}  {\bf  506} 111 
\item Huebner W F, Merts A L, Magee N H and Argo M F 1977 {\it Los Alamos Sci. Lab. Rept. LA-6760 M}
\item Iglesias C A, Rogers F J \& Wilson B G 1987 {\it ApJ}  {\bf  322} 45 
\item Iglesias C A  \& Rogers F J 1996  {\it ApJ}  {\bf  464} 943 
\item Loisel G et al. 2009 {\it HEDP} {\bf 5} 173
\item Loisel G et al. 2011 {\it MNRAS} in preparation
\item Porcherot Q, Gilleron F,  Pain J C and  Blenski T , 2010, to be published in HEDP. 
\item Seaton M J 2005 {\it MNRAS}  {\bf  362} 1
\item Seaton M J \& Badnell N R 2004 {\it MNRAS}  {\bf  354} 457
\item Thompson M et al. 1996 {\it Science} {\bf 272} 1300
\item Turck-Chi{\`e}ze S et al. 1993 {\it  Phys Rep} {\bf 230} 59
\item Turck-Chi{\`e}ze S et al. 2001 {\it  ApJ} {\bf 555} L69
\item Turck-Chi{\`e}ze S et al. 2009 {\it  HEDP} {\bf 5} 132
\item Turck-Chi{\`e}ze S et al. 2010 {\it  Astrophys \& Space Science} in press
\end{thereferences}
\end{document}